\documentclass[aps,prl,twocolumn,floats]{revtex4}

\usepackage{graphicx,psfig}
\usepackage{dcolumn}
\usepackage{bm}

\begin{document}

\bibliographystyle{prsty}

\title{
%
Comment on cond-mat/0406502 by V. Yukalov and E. Yukalova \\ on
Coherent Nuclear Radiation \vspace{-1mm} }

\author{
E. M. Chudnovsky$^1$ and  D. A. Garanin$^2$ }
 \affiliation{ \mbox{$^1$Department of Physics and Astronomy, Lehman College, City
University of New York,} \\ \mbox{250 Bedford Park Boulevard West,
Bronx, New York 10468-1589, U.S.A.}
\\
\mbox{$^2$Institut f\"ur Physik, Johannes-Gutenberg-Universit\"at,
 D-55099 Mainz, Germany}\\}

\begin{abstract}
We show that the argument of Yukalov and Yukalova that
dipole-dipole interaction prevents a system of magnetic dipoles
from exhibiting superradiance unless assisted by a resonator is
incorrect.
\end{abstract}

\maketitle

The e-print of Yukalov and Yukalova \cite{YY} is a review of
theory and experiment on coherent electromagnetic radiation by
nuclear spins. In addition, the authors suggested, in accordance
with our earlier work \cite{CG}, that crystals of molecular
magnets can be sources of coherent microwave radiation that is
essentially stronger than the one from nuclear spins. The purpose
of this Comment is to correct an error that propagates through
several sections of Ref. \onlinecite{YY}. It originates in Section
6.1 that states that in order for the spin superradiance (SR) to
occur, the SR rate, ${\Gamma}_{rad}$, must exceed the rate of
dephasing, ${\Gamma}_2$, due to magnetic dipole interaction
between the spins, electronic or nuclear. The ratio of the two
rates has been computed by the authors of Ref. \onlinecite{YY} as
\begin{equation}\label{ratio}
\frac{{\Gamma}_{rad}}{{\Gamma}_2} = \frac{2}{3n_0}(k_0L)^3\;,
\end{equation}
where $n_0$ is the number of nearest neighbors for each spin,
$k_0$ is the wave vector of the radiation, and $L$ is the size of
the system. According to Yukalov and Yukalova, the condition
${\Gamma}_{rad} \gg {\Gamma}_2$ is in conflict with the other
condition of the SR: $k_0L \ll 1$. This made them to conclude that
the SR based upon magnetic dipoles was impossible unless assisted
by a resonator. While we agree that the use of the resonator may
greatly enhance SR, we disagree with the above argument.

We first notice that Eq.\ (\ref{ratio}) does not distinguish
between the SR based upon magnetic dipoles and the SR based upon
electric dipoles. Thus, if the above argument was correct, it
would invalidate not only the model of non-resonator SR induced by
a field pulse in a crystal of molecular magnets \cite{CG} but also
the conventional Dicke model \cite{Dicke} of SR based upon
electric dipoles. The error stems from the use of
\begin{equation}\label{Gamma-2}
{\Gamma}_2 = n_0{\rho}{\gamma}{\mu},
\end{equation}
where ${\rho}$ is the concentration of spins, ${\gamma}$ is the
gyromagnetic ratio, and ${\mu}$ is the magnetic moment associated
with the spin. Eq.\ (\ref{Gamma-2}) is a frequency equivalent of
the dipolar magnetic field exerted on any given spin by the
neighboring spins. It was introduced by Abragam \cite{Abragam} as
a measure of the NMR linewidth when spins are randomly oriented.
It is that distribution in the orientation of the magnetic dipoles
that contributes to the linewidth. Random orientation is usually
the case for nuclear spins -- they are difficult to align even by
a high field at low temperature due to the smallness of nuclear
magnetic moments. The latter is not true for crystals of molecular
nanomagnets, though.

In our model \cite{CG}, the spins, initially aligned by strong
magnetic field, preserve the phase coherence when the field is
swept in the opposite direction. During the collective relaxation
towards the direction of the field, the mutual orientation of the
spins does not change. This is manifested by the conservation of
the length of Dicke pseudospin. The same is true for the Dicke
model \cite{Dicke}. If the electric dipoles considered by Dicke
had random orientations, as Yukalov and Yukalova assumed for
magnetic dipoles, there would be no Dicke superradiance.

One should also notice that the SR does not disappear when the
size of the system, $L$, grows beyond the wavelength of the
radiation, ${\lambda}_0 = 2 \pi/k_0$. In this limit the intensity
of the radiation simply oscillates on $k_0L$ with amplitude that
slowly goes to zero on increasing $L$ \cite{CG-phonons}. Thus, the
restriction on $k_0L$ is significantly softer than $k_0L \ll 1$
assumed in Ref. \onlinecite{YY}.

The conclusion of Yukalov and Yukalova based upon equations
(\ref{ratio}) and (\ref{Gamma-2}) is an overkill. The
dipole-dipole interaction does not prevent systems of electric or
magnetic dipoles from exhibiting superradiance.


\end{document}